# Kozai Mechanism and real Kozai: Raw Talents of the Highest Order in Mathematics

Aswin Sekhar

Centre for Earth Evolution and Dynamics, Faculty of Mathematics and Natural Sciences, University of Oslo, Norway, E-mail: aswin.sekhar@geo.uio.no

I think it was the first time in a mainstream scientific conference (referring to international conference of Asteroids, Comets, Meteors 2012 held at Niigata, Japan) where I witnessed a really long extended set of claps and applause after a very technical science talk filled with complicated equations and expressions. Not often you get to hear such a long tribute-like applause amongst serious scientists who are usually stingy in these things and prefer a few seconds short standard clap. The whole event (Kozai 2012) in itself was a bit unique because the talk title was 'Kozai Mechanism' and the speaker's name was Kozai. Again, not often scientific speakers get such a privilege in having the title and speaker sharing the same name!

As a student (Sekhar 2014) in celestial mechanics those days, I was aware of this legendary astrophysics figure Yoshihide Kozai from conversations with my Ph.D supervisor Dr David Asher and director Prof Mark Bailey of the observatory where I worked in United Kingdom. His interesting theory named Kozai mechanism (Kozai 1962) led to extremely fascinating implications in the evolution of orbits in both solar system (Bailey, Chambers & Hahn 1992, Asher & Steel 1996, Morbidelli 2011, Werner & Ivanov 2015, Sekhar et al. 2017) and exoplanetary systems (Naoz et al. 2011, 2013, Li et al. 2014, 2015) from a celestial mechanics point of view. Although I have read Kozai's papers and other numerous papers citing Kozai mechanism during that time, I never had the fortune to listen (Kozai 2012) to this legend in person before this conference held at Japan.

There is one remarkable irony I immediately noticed during this gentleman's talk i.e. his way of speaking and conveying the English words (or anything non-mathematical for that matter) in the text on slides during the talk was slightly incoherent and unclear. Furthermore his navigation with laser pointer to slides were totally out of place but as soon as he spoke about the mathematical expressions, equations and technical terms in mathematics, his pronunciation and speech was crystal clear for anyone's ears and the talk suddenly came to full life astounding with mathematical rigour. It gave an impression that trigonometric 'sin' and 'cot' were more familiar to him than conventional 'sin' and 'cot' in English!!! ;) It was clearly evident that internally he was more comfortable in the plane of mathematics on the highest level rather than the conventional speaking language of humans in general. This disparity in his comfort level of handling between the human language and the so called 'god's language' of mathematics was clear and transparent during the talk (at least to my childish ears!). The uniqueness of this talk has lingered in my brain as a fond memory ever since.

Some months back I was listening to Graham Farmelo's (who is a good biographer of Dirac) talk about the life and work (Farmelo 2009) of legendary physicist P A M Dirac, who is well known for unifying pivotal aspects of relativity and quantum mechanics (Dirac 1995), and he was expressing Dirac's inability and lack of interest in expressing long sentences and conversations in English but at the same time, highlighting his exceptional ability in making scientific theories extremely elegant with short mathematical equations. There were even some philosophers and artists who have expressed that Dirac's mathematical equations were like perfect marble statues fallen from the heaven. Dirac was certainly a man who sheerly communicated with mathematics (more than anything else) to scientific colleagues and mathematics was his medium of communication in life. Dirac was known to reply to most questions either with silence or just one or two English words (Farmelo 2009) although he would fill black boards, notebooks and his papers with reels and reels of fine mathematical equations.

Coincidentally, I was watching the recent Hollywood film named 'The Man Who Knew Infinity' about the life and work of eccentric genius Indian mathematician Srinivasa Ramanujan. Both excellent biography (Kanigel 1991) of the same title as well as this latest Hollywood film (Brown 2016) clearly indicate that Ramanujan felt more comfortable and was at ease in the 'godly' language of mathematics rather than conventional Tamil or English for scientific conversations. Most of his conversations, manuscripts and notebooks hardly had any Tamil or English words but was decorated with mathematical symbols, notations, formulaes of every possible combination (Kanigel 1991) extending even to margins of notebooks and letters. Whatever he wanted to express in his life, he felt more comfortable in conveying them through mathematics than any other medium.

Last year when I had the fortune of meeting and talking to Prof Manjul Bhargava, a young Field's medalist (http://www.mathunion.org/general/prizes/2014/prize-citations/) of Indian origin based at Princeton University who was one of the speakers (http://www.abelprize.no/c67534/artikkel/vis.html?tid=67578) during the prestigious Abel prize ceremony (Sekhar 2017) in mathematics, hosted by the Norwegian Academy of Science & Letters and Norwegian Royal Family, held at University of Oslo in 2016. I remember him mentioning interesting mathematical patterns and symmetry about the Indian musical instrument Tabla and waving his hands (again following the pattern of little human language at that time and giving more hand expressions plus mathematical terms) to describe those musical patterns and rhythm (http://www.thehindu.com/opinion/Big-screen-and-the-hero-sum-game/article13990457.ece). There was this innocent streak and gleam of a pristine mathematical genius in those eyes and actions which all seem to make an interesting pattern connecting all the beautiful mathematical minds listed above.

After Prof Kozai's talk, I had the privilege of talking to him for couple of minutes. I remember telling him that I found his paper on Kozai mechanism really interesting but clarified that I could not grasp all the complicated mathematical expressions in his paper in fine detail. He politely replied that he reduced the number of equations in his paper during the writing of the paper so that there is more space for text and explanation! Anyone who has read Kozai's classic papers knows that he has written the manuscript (Kozai 1962) in fine mathematics with minimum English words!!! :) This in turn is an automatic indication of their penchant for equations and the inherent comfort in handling mathematical expressions like their toys!

From Prof Kozai's innocent reply, I got convinced that all great mathematical prodigies think alike irrespective of where they originated from, whether it be Japan or England or India or Canada. It looks like their genes have inherently adopted that their first language will remain as the blessed 'godly language' of mathematics with which they think, write and communicate while every other form of language is just secondary for the menial survival matters of the mortal world!

Beyond the metaphysical beauty and harmony in an Einsteinian sense, it is extremely interesting to note that there is a certain surreal connection of simplicity and symmetry between the finest minds in mathematics beyond borders, boundaries and beliefs. May be that is what unifies mathematicians of the finest order in history on a sublime plane!

May be one day the whole world will get to see an international documentary or Hollywood film about the life and work of raw talents like Hagihara (Hagihara 1957, 1979), Hirayama (Hirayama 1918, 1927) who are other well known celestial mechanics stalwarts from Japan in addition to Kozai who belong to this distinguished category of divine mathematical minds from Japan who prefer to speak the purest language of mathematics of highest order in their august body of celestial work! That might perhaps inspire more and more school children and college students from Japan to pursue and continue the rich Japanese legacy in celestial mechanics adorned by these mathematical gems in orbits!

**References:**

Asher D. J., Steel D. I., 1996, MNRAS, 280, 1201.

Bailey M.E., Chambers J.E., Hahn G., 1992, A&A, 257, 315.

Brown M. 2016, The Man Who Knew Infinity, Warner Bros. UK.

Dirac P. A. M. 1995, The Collected Works of Dirac 1924-1948, Cambridge; New York: Cambridge University Press, c1995, edited by Dalitz, Richard H.

Farmelo G. 2009, The Strangest Man: The Hidden Life of Paul Dirac, Quantum Genius, Faber and Faber, United Kingdom, ISBN 978-0-571-22278-0.

Hagihara Y. 1957, Stability in Celestial Mechanics, Tokyo: Kasei Publication.

Hagihara Y. 1979, Dynamics of the Solar System; Proceedings of the Symposium, Tokyo, Japan, May 23-26, 1978. Edited by R. L. Duncombe. Symposium sponsored by the IAU, COSPAR, IUTAM, and Japan Society for the Promotion of Science Dordrecht, D. Reidel Publishing Co. (IAU Symposium, No. 81), 1979., p.1.

Hirayama K. 1918, Astronomical Journal, vol. 31, iss. 743, p. 185-188.

Hirayama K. 1927, Japanese Journal of Astronomy and Geophysics, Vol. 5, p.137.

Kanigel R. 1991, The Man Who Knew Infinity, C. Scribner's, USA, ISBN 978-0-684-19259-8.

Kozai Y. 2012, Asteroids, Comets, Meteors 2012, Proceedings of the conference held May 16-20, 2012 in Niigata, Japan. LPI Contribution No. 1667, ID 6195.